\documentclass[sigconf,natbib=false]{acmart}

\AtBeginDocument{%
  }

\setcopyright{acmlicensed}
\copyrightyear{2018}
\acmYear{2018}
\acmDOI{XXXXXXX.XXXXXXX}

\RequirePackage[
  datamodel=acmdatamodel,
  style=acmnumeric,
  ]{biblatex}

\begin{document}

\title{LLMs in HCI Data Work: Bridging the Gap Between Information Retrieval and Responsible Research Practices}

\author{Neda Taghizadeh Serajeh}
\email{nedath1378@gmail.com}
\orcid{0009-0001-0849-4426}
\affiliation{%
  \institution{Sharif University of Technology}
  \streetaddress{P.O. Box 1212}
  \city{Tehran}
  \state{Tehran}
  \country{Iran}
}

\author{Iman Mohammadi}
\email{imanmohammadi@sharif.edu}
\affiliation{%
  \institution{Sharif University of Technology}
  \city{Tehran}
  \state{Tehran}
  \country{Iran}
}

\author{Vittorio Fuccella}
\email{vfuccella@unisa.it}
\affiliation{%
  \institution{University of Salerno}
  \city{Fisciano (SA)}
  \country{Italy}
}

\author{Mattia De Rosa}
\email{matderosa@unisa.it}
\affiliation{%
  \institution{University of Salerno}
  \city{Fisciano (SA)}
  \country{Italy}
}

\renewcommand{\shortauthors}{Trovato et al.}

\begin{abstract}
  Efficient and accurate information extraction from scientific papers is significant in the rapidly developing human-computer interaction research in the literature review process. Our paper introduces and analyses a new information retrieval system using state-of-the-art Large Language Models (LLMs) in combination with structured text analysis techniques to extract experimental data from HCI literature, emphasizing key elements. Then We analyse the challenges and risks of using LLMs in the world of research. We performed a comprehensive analysis on our conducted dataset, which contained the specified information of 300 CHI 2020-2022 papers, to evaluate the performance of the two large language models, GPT-3.5 (text-davinci-003) and Llama-2-70b, paired with structured text analysis techniques. The GPT-3.5 model gains an accuracy of 58\% and a mean absolute error of 7.00. In contrast, the Llama2 model indicates an accuracy of 56\% with a mean absolute error of 7.63. The ability to answer questions was also included in the system in order to work with streamlined data. By evaluating the risks and opportunities presented by LLMs, our work contributes to the ongoing dialogue on establishing methodological validity and ethical guidelines for LLM use in HCI data work.
\end{abstract}





\maketitle

\section{Introduction}
The field of human-computer interaction (HCI) has been developing rapidly these years, and the wide range of information and its scientific contributions indicate proof of that statement. As the HCI knowledge extends exponentially, the need to collect efficient and accurate information for the literature review part of research becomes critical\cite{ModernInformationRetrieval}.
In response to this data-rich era, our research tries to introduce a novel approach for HCI researchers to extract and interpret information from scientific documents. Our study is a new method of information retrieval by using advanced techniques and Large Language Models (LLM) combined\cite{proximity}. By experimenting with the capabilities of prominent LLMs such as GPT-3.5 \cite{chatgpt_chatgpt} and Llama2 \cite{Llama_Llama}, we can reach our goal, which is to improve HCI research by extracting accurate and deep understanding of experimental data and answering user questions from the scientific papers\cite{kamalloo2023hagrid}.
The reason we chose these experimental data as the recording parameters of each paper is that the complexity of extracting this type of data indicated high\cite{ModernInformationRetrieval}, and it comes from different understandings researchers may have from them. However, because of this type of complexity, we aim to extract this valuable information with high accuracy from each paper\cite{ModernInformationRetrieval}. Moreover, some of the parameters extracted by the models were very different from the obtained human parameters, which we investigated.
We also review a comprehensive analysis performed on a smaller selected dataset, randomly, consisting of 300 papers from CHI conferences between 2020 and 2022. Our methodology not only has good accuracy in data mining but also addresses the complexities of user studies, interviews, and controlled experiments. Also, due to our manual checks, we recorded the values of each required parameter specifically. Eventually we were able to report the accuracy of our work, and we reached a favorable result\cite{jeronymo2023inparsv2}.
Another feature of our approach is the ability to process a wide range of file formats. This adaptability makes our method flexible, allowing it to keep up with the changing world of digital publishing. Furthermore, users can seamlessly engage with our developed user-friendly interface, finding answers to their questions on the input documents.

\section{Related Works}

The field of Human-Computer Interaction (HCI) and scientific literature information extraction has seen the appearance of numerous innovative tools designed for specific purposes. Before discussing other similar tools in this domain, it is crucial to mention our system. Our system is meticulously designed to combine the capabilities of Large Language Models with structured text analysis. By doing so, we’re not just merely extracting data but interpreting, categorizing, and presenting the specified information in a coherent and user-friendly format. Our approach involves conducting a literature review focused on the experimental sections of HCI papers. This section will provide a comprehensive overview of several notable tools, explaining their functionalities and contextualizing them within HCI and information extraction. Throughout the exploration of each tool, we will highlight their advantages and compare them to our system.

\subsection{Existing Tools}

This section explores different information extraction tools, focusing on their features and functionalities. Also, in Table 1, we provide a brief view of these tools and our work.

\subsubsection{Grobid}

Grobid \cite{GrobidCombiningAutomatic} is a machine-learning library that extracts structured metadata from academic documents\cite{grobid}, focusing on PDF extraction. It excels in bibliographic database applications, but our tool offers a more general approach, despite longer processing times using LLMs.

\subsubsection{CoreNLP}

Stanford CoreNLP, developed by the Stanford NLP Group\cite{corenlp}, offers a variety of linguistic tools for tasks like tokenization and parsing across multiple languages. Our system combines preprocessing and direct data extraction, providing users with the flexibility to choose between streamlined but slower information extraction or faster offline preprocessing based on their needs.

\subsubsection{Elastic Search}

Elasticsearch is a powerful search engine that analyzes big data in real-time\cite{elasticsearch}. It uses flexible JSON documents and a compatible Web API. The system discussed improves search by extracting and interpreting data from scientific literature, especially in HCI. This integrated approach offers users structured information, giving the tool a unique edge in data processing.

\subsection{LLMs}

As the digital age progresses, Large Language Models (LLMs) have come to the forefront of information retrieval\cite{zhu2023large}, representing a paradigm shift in how data is parsed, understood, and presented\cite{ram2023incontext}. Rooted in deep learning, LLMs like GPT-2\cite{radford2019language, chatgpt_chatgpt} have the capacity to comprehend vast swaths of text\cite{tan2019lxmert}, extracting not just explicit facts but also nuanced insights\cite{jeronymo2023inparsv2}. They are also adaptive\cite{manathunga2023retrieval}, learning from new data and user interactions, making them adept at handling the dynamism of modern digital content\cite{tan2023reimagining}. Moreover, with their capability to work across languages and domains, they transcend geographical and disciplinary boundaries\cite{ModernInformationRetrieval}.

\begin{table*}[t]
\centering\def\arraystretch{1}
\centering
\footnotesize{
\setlength{\tabcolsep}{7.7pt}%
\caption{Comparison of Various Systems.}
\begin{tabular}{lllllll}
\cmidrule[1.5pt]{1-7}
Name & Main Purpose & Programming Language & Open Source & License & Speed & Scalability\\
\bottomrule
Grobid & PDF extraction & Java & Yes & Apache 2.0 & Over 36 PDFs per second & Moderate \\
CoreNLP & NLP Toolkit & Java & Yes & GPL & Over 1000 words per second & High \\
Elasticsearch & Search and Analytics Engine & Java & Yes & Apache 2.0 & Over 1000 documents per second & High \\
Our System & Information Extraction & Python & Yes & MIT & It depends on the API model used & High \\
\cmidrule[1.5pt]{1-7}
\end{tabular}}
\label{tab:system_comparison}
\end{table*}

\section{Method}

Our project aims to develop a tool that automates the extraction of information from empirical papers in the field of Human-Computer Interaction, following the APA Style guidelines.

The experiment aims to gather important information for HCI research papers, including details like the number of participants, recruitment methods, tasks, type of experiment, and specifics about experimental variables and trials. This data is valuable for literature reviews in the field.

In order to improve text processing for model analysis, we used techniques like Named Entity Recognition (NER) to identify important entities and Keyword Extraction to prioritize key phrases. We also removed unnecessary elements like references and headers to minimize program overhead and focus on relevant content. Our system supports HTML and PDF formats commonly used in academic publishing. HTML is great for online reading, while PDFs are convenient for offline access. We also handle compressed files like zip, rar, and 7z, making it easier to work with academic documents. The document processing uses a Python framework with BeautifulSoup for HTML, PyPDF2 for PDF, and LLMs for OpenAI and Llama2 API integration to automatically extract important experimental data from scientific papers, making data handling faster and more accurate. The implementation of our information extraction system faced challenges with HTML and PDF formats. HTML changes in tags, classes, and styles required adapting the parser to each file's unique structure. PDFs, with binary formats and embedded elements, posed obstacles in data extraction. Addressing these complexities was crucial for ensuring the reliability of our system. 

The main focus of this implementation is the interaction with Llama2 \cite{Llama_Llama} and GPT3.5 \cite{chatgpt_chatgpt} APIs, where the text of each article is processed based on a defined prompt. Implemented a method that acts as a bridge, sending paper content to an API and retrieving targeted details. The code includes features to handle errors, switch between API keys, and measure performance metrics like latency and memory consumption for extraction efficiency insights.

The prompt discussed the need to include specific output values related to a parameter model, such as Number of Participants, Recruitment Method, Number of Tasks, Type of Experiment, Experimental Variables, and Number of Trials. It also highlighted the importance of avoiding redundant or irrelevant explanations in the command. Emphasis was placed on key items to improve understanding. The suggestion was made to focus on writing a parameter model with a comprehensive explanation, summary, and emphasis on crucial parts, particularly for long short-term memory (LSTM) models.

We assessed the system's performance using measures like latency, processing speed, and memory consumption, summarized in Table 2. Latency, the time to process a document, is crucial for responsiveness. Lower latency enables handling large datasets quickly. Processing speed, measured as papers processed per unit time, gauges system consistency across diverse datasets. Memory consumption indicates resource utilization, ensuring scalability for larger inputs without slowdowns or crashes.

\subsection{Dataset Construction}

We collected 2148 articles from 2020 to 2022 CHI conferences to study HCI research trends. We randomly picked 300 articles (100 per year) and reviewed them with our system. Each article was carefully annotated based on specific criteria.

\subsubsection{Number of participants and recruitment method}
The dataset shows how participants were chosen for each study, considering factors like age, gender, and relevant skills.

\subsubsection{Number of tasks}
We noted how many tasks participants had in each study, giving us insights into workload and testing complexity.

\subsubsection{Type of Experiment}
We sorted studies by their experimental design, covering methods like user studies, interviews, lab experiments, and online surveys. Brief explanations were given for the chosen experimental approaches.

\subsubsection{Experimental variables}
Detailed annotations were made to identify and specify the experimental variables used in each study. This comprehensive documentation includes the independent variables, their associated levels, and any control variables that are integrated to reduce potential confounding effects.

\subsubsection{Number of trials}
For experiments involving repeated or repetitive tasks, we detailed the total number of trials performed by participants for each specific task. These data shed light on the participation and repetition of participants in the experiments.

Considering these criteria, our dataset not only provides a snapshot of recent HCI research methods but also serves as a good resource for understanding and comparing the complexities of experimental design across many studies. This dataset serves as a valuable asset for researchers seeking to explore the landscape of HCI research methods and trends.

\section{Evaluation}

Two evaluation methods, mean absolute error (MAE) and accuracy, are used for evaluating the system's performance by comparing the accuracy and similarity of information extracted by the system to manually selected vectors for each paper.

\subsection{MAE}

Our approach involves representing each article as a vector containing extracted features. Each vector serves as a representation of the corresponding paper. The evaluation process is focused on evaluating the alignment of the output vectors of the models with the manually generated vectors for each article. Specifically, we use the MAE method and treat the problem as a regression task. The goal is for the models to accurately reconstruct the vector of each article. MAE represents the regression error, which indicates the closeness of the reconstructed vectors by the model to the original vectors.
For the GPT-3.5 model \cite{chatgpt_chatgpt}, LOSS MAE was equal to 7, which indicates a relatively strong performance in reconstructing paper vectors. In comparison, the Llama model \cite{Llama_Llama} achieved a LOSS MAE of 7.63, indicating a slightly lower accuracy. Therefore, the GPT-3.5 model \cite{chatgpt_chatgpt} performed better than Llama \cite{Llama_Llama} in this evaluation method.
 
$$ 
\text{MAE} = \frac{1}{n} \sum_{i=1}^{n} I_{|y_i - \hat{y}_i| \leq \text{approximation level}}
$$

MAE represents the Mean Absolute Error.

\(n\) is the total number of data points (articles)

\(y_i\) represents the actual value (manually obtained vector for each article)

\(\hat{y}_i\) represents the predicted value (vector generated by the model)

"Approximation Level" is the threshold for the absolute difference that determines whether the indicator function $I$ equals 1 or 0

\(|y_i - \hat{y}_i| \leq \text{approximation level}\) is the indicator function that checks whether the absolute difference between \(y_i\) and \(\hat{y}_i\) is less than or equal to the specified approximation level.
 
\subsection{Accuracy}

In this evaluation method, we look at the features extracted from articles as labels. We compare the labels generated by the system with manually collected ones to measure accuracy. Higher accuracy means more precise feature extraction. Overall, GPT-3.5\cite{chatgpt_chatgpt} performed slightly better in predicting features compared to manual evaluation.

\subsubsection{Approximation of one}

For minor numerical discrepancies, a tolerance of one is allowed, enhancing accuracy calculations. Under this criterion, GPT-3.5's accuracy for numerical features is 62\%, surpassing Llama's 59\%. This emphasizes the data retrieval system's efficiency.

\subsubsection{Baseline approach} 
A baseline comparison using randomly generated samples from a fitted normal distribution against the actual data reveals a mere 13\% accuracy. This underscores the substantial advancement of our models over random approximation, with GPT-3.5 and Llama outperforming the baseline by 45\% and 43\%, respectively.

\section{Result}

In the early stages of this project, our goal was to implement an efficient system for information retrieval from scientific articles, coupled with an interface that users found effortless to use. Then implementing LLM information retrieval as the main method was a strategic decision driven by its proven effectiveness in handling large volumes of academic data and its adaptability to diverse research contexts. Bearing these ideas, a comprehensive analysis was made from the two models, GPT \cite{chatgpt_chatgpt} and Llama \cite{Llama_Llama}. GPT \cite{chatgpt_chatgpt} consistently outperformed Llama \cite{Llama_Llama} across various accuracy metrics. While both models had high capabilities, GPT’s \cite{chatgpt_chatgpt} baseline accuracy (58\%) marginally surpassed Llama’s (56\%) \cite{Llama_Llama}. This difference was further accentuated when assessing the Mean Absolute Error (MAE), where GPT's \cite{chatgpt_chatgpt}predictions showed a lower deviation from actual values with an MAE of 7, compared to Llama’s 7.63 \cite{Llama_Llama}. To specifically determine the GPT \cite{chatgpt_chatgpt} and Llama models’ \cite{Llama_Llama} efficacy, we evaluated some more performance metrics in specific tasks. Table 1 shows that when it comes to processing speed, GPT \cite{chatgpt_chatgpt} is about 7 times higher than Llama \cite{Llama_Llama}. The max token limit was another limitation that we had to face in the system’s implementation. Both of the models had the same amount of tokens limit while using their APIs. In terms of latency, GPT 3.5 \cite{chatgpt_chatgpt} exhibited a significantly faster response time of approximately 4.52 seconds per paper, while Llama \cite{Llama_Llama} took notably longer with an average latency of 31.22 seconds per paper. When evaluating memory consumption, GPT 3.5 \cite{chatgpt_chatgpt} utilized around 166.42 MB of memory, in contrast to the more efficient Llama \cite{Llama_Llama}, which consumed just about 91.70 MB.

\begin{table}[t]
\centering
\caption{Comparative Performance Metrics of LLMs.}
\setlength{\tabcolsep}{5pt} % Adjusted for single column width, might need fine-tuning
\footnotesize{
\begin{tabular}{lll}
\cmidrule[1.5pt]{1-3}
Parameter/Model & GPT 3.5 & Llama2 \\
\midrule
Accuracy & 58 & 56 \\
mean absolute error & 7 & 7.63 \\
Processing Speed (papers/sec) & 0.2210592478 & 0.03103395873 \\
Max Token Limit & 4,096 tokens & 4096 tokens \\
Latency (s) & 4.5236741273 seconds & 31.2227663099 \\
Memory Consumption (MB) & 166.4164453125 & 91.702734375 \\
\cmidrule[1.5pt]{1-3}
\end{tabular}
}
\label{tab:performance_comparison}
\end{table}

\section{Discussion}

The paper highlights the advantages of using LLMs in a data recovery tool. These models improve accuracy, efficiency, and user understanding of the recovered information. The tool can handle various file formats, including PDF, HTML, 7z, zip, and rar.
Another advantage of our study is the collection and use of a comprehensive dataset of 2148 CHI articles related to human-computer interaction (HCI). We selected a subset of 300 articles underwent a rigorous human review process, which allowed us to measure the accuracy of our methodology using the exact recall method. Careful examination of this subset ensures that it has the necessary diversity and specificity, making it particularly valuable for quality assessment. Therefore, given the scope and limitations of our study, focusing on a smaller yet representative sample was a more manageable and efficient approach. This strategy not only ensured accuracy but also reduced surveillance risk. In doing so, we maintained the integrity of our research while increasing efficiency, given the study's specific goals and limitations.

Almost all the models that are prepared and taught today can meet the needs of users with accuracy and information, but the important thing is to choose them according to your needs. We used 2 models, GPT-3.5 \cite{chatgpt_chatgpt} and Lama2, for this system.
The GPT-3.5 \cite{chatgpt_chatgpt} model has unparalleled accuracy because it has been trained on about 45 terabytes of data and 175 bytes of trained parameters.
Llama Model 2 \cite{Llama_Llama} is open source and promotes continuous improvement from the global developer community. Meta trained and published Llama-2 \cite{Llama_Llama} in three model sizes: 7, 13, and 70 billion parameters, and they are trained on 2 trillion tokens.

\subsection{Reasons of Accuracy Decrease}
In the process of manual verification and comparison of our system's outputs with the original data from the articles, several noteworthy observations came to our mind. These insights provide valuable context for interpreting the performance of the models.

Handling of Multi-Stage Experiments: 
In multi-stage experiments, a common issue is that models often only report the number of participants in the initial phase, ignoring those in later stages. This can be problematic when participant counts vary across stages. Despite this limitation, manual checks were conducted to address the issue.

Task Counting Accuracy: 
Another noteworthy observation concerns the extraction of task counts. When articles state, for example, "We administered tests to participants across n tests, including m phases," the models often extract the value n, whereas our desired output was $n \times m$. This highlights the need for refining the models' ability to capture the intricacies of task counting within sentences.

These observations shed light on specific challenges and nuances encountered during the evaluation process. 

\subsection{Limitation}

While our study provides a significant contribution to information mining in human-computer interaction (HCI) research, it is important to acknowledge its inherent limitations, which include:

Dependency format.
The effectiveness of our system depends on the compatibility of the input data with the specified file formats (e.g. text, pdf, HTML, zip, rar, 7z).

Limited data set selection.
Although we selected a representative subset of articles for analysis, our dataset includes a specific time frame (CHI articles from 2020 to 2022). Also, finding the required information and checking it manually is very time-consuming and difficult, and because some of the data collected to calculate the accuracy of the system is collected by humans, it is definitely not 100\% error-free.

Dependence on online access.
Our data retrieval tool relies on online access to large language models (LLM) through APIs. As a result, it cannot be used offline.

Experimenting based on complex parameters.
In the experiment part, we assessed our system’s proficiency in extracting specific values from documents by finding the values of the Number of Participants and Recruitment Method, Number of Tasks, Type of Experiment, Experimental Variables, and Number of Trials of 300 papers in our dataset. These parameters are notably ambiguous and challenging to pinpoint in scientific papers which contain experimental data. This complexity comes from the fact that each can have different definitions depending on the research domain. 

While these limitations are inherent to our current implementation, they also present opportunities for future research and tool modification to address these challenges and expand the usability and utility of the tool.

\section{Future Work}

This research improves how computers and humans interact, paving the way for future improvements. There are suggestions for making the work even better.
By modifying the learning of  LLMs with a specialized focus on the terminology and context of HCI. Also, giving more detailed information about the type of parameters and how to extract them efficiently.
Using the Google Bard model for real-time information extraction.
Considering offline operation to enhance performance and reduce costs in LLM-based data retrieval.

\printbibliography

\end{document}